\documentclass[aps,prl,twocolumn,showpacs,a4paper,amsmath,amssymb,byrevtex,floatfix]{revtex4-1}

\usepackage[latin1]{inputenc}
\usepackage{graphicx}
\usepackage{dcolumn}
\usepackage{bm}
\usepackage{times}
\usepackage{color}

\newcommand\varpm{\mathbin{\vcenter{\hbox{%
	\oalign{\hfil$\scriptstyle+$\hfil\cr
	\noalign{\kern-.3ex}$\scriptscriptstyle({-})$\cr}%
}}}}

\begin{document}

\title{Electronics without bridging components}

\author{V. M. Garc\'{\i}a-Su\'arez}
\affiliation{Departamento de F\'{\i}sica, Universidad de Oviedo \& CINN, 33007 Oviedo, Spain}
\email{vm.garcia@cinn.es}

\date{\today}

\begin{abstract}
We propose a new paradigm of electronic devices based only on two electrodes separated by a gap, i.e. without any functional element bridging them. We use a tight-binding model to show that, depending on the type of material of the electrodes and its structure, several electronic functionalities can be achieved: ohmic behaviour, rectification, negative differential resistance, spin-filtering and magnetoresistance. In particular, we show that it is possible to deliver a given functionality by changing the coupling between the surface and bulk states and between the surface states across the gap, which dramatically changes the current-voltage characteristics. These results prove that it is possible to have functional electronic and spintronic elements on the nanoscale without having physical components bridging the electrodes. 
\end{abstract}


\maketitle

\section{Introduction}

As the size of the electronic components approaches the atomic limit, where typical complementary metal-oxide-semiconductor (CMOS) elements start to fail, it is necessary to look for functional alternatives. A series of nanoscale electronic systems have already been proposed, which include nanometric elements such as molecules \cite{MccCM04}, nanotubes \cite{AvoPIEEE03}, nanowires \cite{PalN12}, and the most recent two-dimensional (2D) materials \cite{OsaAM12}. These systems would allow to go beyond typical silicon-based components \cite{ITRS15} and develop much faster and reliable electronic systems. The feasibility of many of these new elements has already been demonstrated in lots of experiments and different electronic functionalities have already been proven to work, such as rectification \cite{NguIEEETED89}, switching \cite{JoaCPL97}, negative differential resistance (NDR) \cite{LyoS89,CheAPL00}, spin filtering \cite{GohS211} and magnetoresistance \cite{TsuN99,GarPRL99}. However, the development of these electronic components is still in their infancy and no widely-used technological application has been reached so far. In addition, problems due to atomic migration, defects and quantum tunneling render these elements useless. For example, in systems based on molecules or similar nanometric elements \cite{LorNN13}, due to the movement of the coupling groups or the atoms of the surface, measurements of the transport properties show a rather large variability and lack of reproducibility, which in turn translate into a lot of noise and make such systems non-viable. These problems need therefore to be solved in order to keep the ongoing miniaturization of electronic components, i.e. keep alive Moore's law \cite{MorE65,ITRS15}, and, at the same time, make room for the development of ``More than Moore'' approaches \cite{Top15Mo}.

Instabilities due to the electrodes can be removed in principle by using covalently bonded materials, which have a small number of defects and whose directional bonds prevent the migration of atoms. In recent years it has been proposed the use of materials such as graphene \cite{KolPRL07,PriNL11} or other 2D systems. The idea is to keep two sheets of the 2D material separated by a nanogap and bridge them with a nanoscale element, such as a molecule, with large coupling groups \cite{MarJACS08}. The large and planar coupling groups would average all possible contact configurations between them and the 2D surface and would give in principle stable and reproducible transport properties. This technology still poses, however, a series of problems due to the mobility of the molecule on the surface, which easily slides due to the weak van der Waals coupling. Also, the mismatch between the molecular states and the Fermi level \cite{GarPRB13} and the fact that the molecule has to be relatively large to effectively connect both electrodes, translates into a conductance that is rather small and decreases exponentially with the length of the molecule. Such problems make these systems at present not amenable to functioning as electronic components.

Most of the problems, if not all, could be circumvented by getting rid of the functional element bridging both electrodes, which is in general the main source of noise through the coupling to the leads. The transport properties would be then more reliable, stable and reproducible. This can be accomplished through the use of surface \cite{GarN10,XiuNN11,NirNL18} or graphene edge states \cite{KatJVSTB17}, which have already been proposed for electronics. We advocate in this article, however, to go a step further and use the presence or absence of localized states to tailor the surfaces or edges and get a desired functionality. This can be achieved by using systems ranging from metallic surfaces with adsorbates to edges of 2D materials with different shapes, which are specially stable. Even though this looks like a radical and non-viable simplification, which would wipe out any electronic functionality --encapsulated in the molecular backbone--, we prove in this work that electronic systems with only two electrodes separated by a nanoscale gap (surfaces of 3D materials, edges of 2D materials or  terminating ends of 1D materials) can produce a plethora of electronic functionalities that can be used to design different electronic components: resistances, rectifiers, NDR devices (digital circuits, oscillators, amplifiers) and spintronic devices. These new designs should become possible thanks to recent developments in the fabrication of nanoscale materials, which allow to fabricate electrodes with very small separations between them. The nanometric size and separation are specially relevant, because quantum effects are averaged out as the size of the system increases and no relevant functionality emerges for relatively large sizes (for instance, the presence of different defects or impurities in large-area surfaces or long edges could render the transport diffusive instead of ballistic, which could wipe out possible functionalities).

The origin of the electronic functionalities in these systems can be traced back to the shape and/or local composition of the surface or edge on both sides of the gap. Certain impurities/shapes produce localized states that couple to the bulk states and to each other across the gap. This is special significant for 2D materials such as graphene \cite{GarNs18}, but it is not limited to them. Typical surfaces with adatoms, impurities or protrusions or other nanoscale systems terminated in localized states could also give rise to similar phenomena \cite{LyoS89,NguIEEETED89}. These states generate resonances in the zero-bias transmission and lead to effects such as rectification or NDR when a bias is applied. Also, by introducing spin-polarized states such as those present in magnetic elements or zigzag graphene edges, it is possible to generate spintronic phenomena such as magnetoresistance or spin-filtering effects. 

These systems have, on one hand, the disadvantage that the transport is in the tunneling regime and therefore the currents are expected to be small, at least for large enough gaps. However, with recent developments in the fabrication of graphene nanogaps with electro-burning techniques \cite{PriNL11} or with the mechanically controlled break junction technique \cite{CanNN18}, it is possible to make nanometer-wide gaps with sizeable currents that give a clear signal. On the other hand, even though designing surfaces or edges with tailored impurities and shapes is still not possible, recent advances in the fabrication of nanoscale gaps could make it achievable in the near future. 

\section{Methods}

\begin{figure}
\includegraphics[width=\columnwidth]{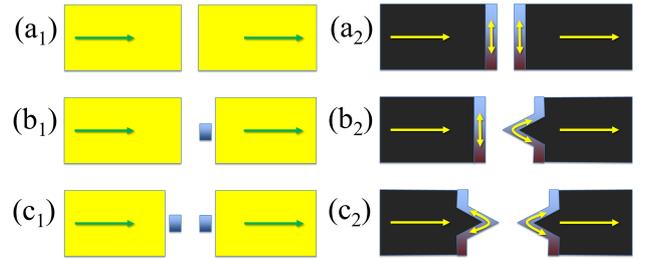}
\caption{\label{Geom} Schematic representation of the three types of systems considered: BB, which can be either two flat surfaces (a$_1$) or two straight edges (a$_2$); BL, which can be either a flat surface and a surface with an adsorbate (b$_1$) or a straight edge and a wedge (b$_2$); and LL, which can be either two surfaces with adsorbates (c$_1$) or two wedges (c$_2$).}
\end{figure}

{\em General scheme}.$\--$ A schematic representation of the systems that we will study is summarized in Fig. (\ref{Geom}), where we show two possible types of electrodes. The left column corresponds to two typical metallic surfaces that can have a localized state in an adatom, impurity or protrusion. The right column represents two electrodes of a 2D material with or without a wedge. In both cases the transport properties depend dramatically on the type of impurity or geometry. In particular, the embedding of the impurity or the orientation of the edge states relative to the transport direction determine their coupling to the bulk transport channels. For example, when the edge is straight and therefore perpendicular to the transport direction, the group velocity of the electrons in these states is perpendicular to the group velocity of the bulk electrons and therefore these do not couple nor contribute to the transmission across the gap. This would be equivalent to having two pristine surfaces in the left column of Fig. (\ref{Geom}). However, when the edge has some imperfection, such as a protrusion or wedge, the group velocity of the edge states acquires a component along the transport direction and contributes to the transmission. In this last case, the edge states couple to the bulk states and influence the transport properties. This would be equivalent to having a surface with an impurity or protrusion in the left column.

The key point is the clear distinction between states that are well coupled to the bulk continuum and states that are localized at the surface/edge. The former tend to produce featureless transmissions, while the latter give rise to transmission resonances. For this reason, we call the configurations with clean surfaces/straight edges "bulk" (B) and the configurations with impurities/wedges "localized" (L). This distinction should not be taken literally because there might be e.g. surface states in flat surfaces that could couple to the bulk continuum and also across the gap, but will serve as a guide to distinguish different transport regimes. We therefore differentiate three possible cases that give rise to different transport properties: two clean surfaces or straight edges (BB), two surfaces with impurities or two wedges (LL) and a combination of the previous two cases (a clean surface or straight edge and a surface with an impurity or a wedge, BL). With these three configurations, which correspond to the three rows of Fig. (\ref{Geom}), it is possible to understand the transport behavior of any two surfaces or 2D layers, no matter how complicated the real configuration in an experiment may be. The reason is that only the (two) closest features would give rise to sizeable currents due to the exponential decay with the separation. The impurities or wedges can also be magnetic --in the following we will consider only magnetic configurations, which are the most general cases-, even though this is not a necessary condition to reproduce most of the functionalities (excluding spintronic effects).

{\em Model}.$\--$ The model is a tight-binding parametrization of the (bulk) states close to the border and the states localized at the surface in different situations. This is a generalization of simple models used before to describe transport in molecular junctions \cite{MarNt09,SpaPRB11} or between graphene edges \cite{CarPRB12,GarNs18}. Even though it is a one-dimensional model, it captures the essential features of more complex systems and produces results that agree with more involved ab-initio simulations. Notice, however, that it might not apply to strongly correlated systems with localized states and large intra-atomic repulsion $U$ (i.e. d or f states), where it would be necessary to go beyond a single particle description. The Hamiltonian describes two electrodes coupled through a vacuum gap, $\hat H=\hat H_\textrm{l}+\hat H_\textrm{r}+ \hat V_\textrm{lr}$, where 'l' and 'r' stand for left and right. In order to properly account for the electronic structure of the surfaces or edges, it is necessary to distinguish between the bulk states and those localized at the surface/edge. The former are delocalized through the whole bulk material, while the latter are localized at the impurity or edge and/or run parallel to it, i.e. their group velocity goes along the direction of the surface/edge and is perpendicular to the flow of electrons, which uncouples them from the bulk states. The contribution of each state in any of the two layers is represented in a second-quantized way as follows:

\begin{equation}
\begin{split}
\hat H_\textrm{l(r)}=&\sum_{\left<i,j\right>;\sigma}\left(t_{ij\sigma}\varpm\delta_{ij}\mathrm{e}V/2\right)\hat c_{i\sigma}^\dagger\hat c_{j\sigma}+\\ +&\sum_\sigma\left(\epsilon_\sigma\varpm\mathrm{e}V/2\right)\hat d_\sigma^\dagger\hat d_\sigma+ \sum_\sigma t_{\textrm{l(r),1d}\sigma}(\hat c_{1\sigma}^\dagger\hat d_\sigma+ \mathrm{h. c.})
\end{split}
\end{equation}

\noindent where $\{\hat c^\dagger\}$ are fermion operators that create electrons in the inner/bulk sites (i.e. those that are not in contact with the surface/edge), $\{\hat d^\dagger\}$ are fermion operators that create electrons in the surface/edge states and $\sigma=\uparrow,\downarrow$ is the spin. The indexes $\{i,j\}$ run to the bulk states of each electrode and the sum $\left<i,j\right>$ goes only to first nearest neighbours. For the on-site and coupling elements, we set $t_{ii\sigma}=\varepsilon_i\equiv\varepsilon$ for $i>1$ (see below) and $t_{ij\sigma}=t$, respectively. The term $\delta_{ij}$ is introduced to reflect the fact that the bias potential is constant on each electrode and therefore acts only on the on-site elements (the basis set is orthogonal). The bulk states in this case are non-magnetic and their on-site energy $\varepsilon$ is placed at zero energy to maintain electron-hole symmetry. The bulk density of states (DOS) is assumed to be constant (wide-band approximation), which is justified in cases where the DOS is smooth around the Fermi level and/or the surface/edge states give the largest contribution to the transmission. Care should be taken, however, in cases with complex DOS, such as e.g. transition metals with d orbitals at the Fermi level.

The surface/edge states have a large magnetic splitting and a electron-hole asymmetry, i.e. $|\epsilon_\uparrow|\neq |\epsilon_\downarrow|$. These states can be uncoupled or weakly coupled to the bulk states. This coupling in real materials can be tuned in principle by using different adsorbate/surface combinations and monoatomic layers or similar systems that change the electronic structure of the surface and its coupling to the bulk, and, in case of 2D materials, by changing the size and/or shape of the wedge. Due to the presence of the surface or edge it is also necessary to introduce corrections in the electronic bulk states close to it. In particular, and for the spin-polarized case, the site that is closest to the edge on each layer has an on-site energy that is slightly different than that of the bulk and develops also a small magnetic splitting due to the interaction with the edge state, i.e. $t_{11\uparrow}\neq t_{11\downarrow}\neq\varepsilon$. This is necessary to properly reproduce the transport properties of straight magnetic edges, as we will show later. Notice again that this particular magnetic configuration is not necessary for this model to work, i.e. similar transport properties can be obtained with non-magnetic surface/edge states (excluding of course spintronic effects). The coupling across the gap is given by the term  $\hat V_\textrm{lr}=\sum_\sigma[\gamma_{\textrm{dd}\sigma}\hat d_{\textrm{l},\sigma}^\dagger\hat d_{\textrm{r},\sigma}+\sum_{i,j=1}^2(\gamma_{\textrm{d}i\sigma}\hat d_{\textrm{l},\sigma}^\dagger\hat c_{\textrm{r},i\sigma}+\gamma_{i\textrm{d}\sigma}\hat c_{\textrm{l},i\sigma}^\dagger\hat d_{\textrm{r},\sigma}+\gamma_{ij\sigma}\hat c_{\textrm{l},i\sigma}^\dagger\hat c_{\textrm{r},j\sigma})+\mathrm{h. c.}]$, which includes the coupling between the localized states and/or the bulk states closest to the surface. In this case, the first index in the coupling terms $\gamma$ ($d$ or $i$) corresponds to a state in the left part and the second index ($d$ or $i/j$) to a state in the right part. The presence of different couplings across the gap gives rise to interference phenomena that can shape the form of the transmission \cite{SpaPRB11}. These couplings can be tuned in real systems by carefully adjusting the distance between both sides of the gap with, for instance, mechanically controllable break junction devices \cite{CanNN18}.

The potential is included as a rigid shift of all the on-site levels of each electrode. This is not an approximation but a rigorous implementation, since the potential falls mainly in the gap, which gives the highest resistance. The physical separation between electrodes and the absence of elements bridging them eliminates also the possibility of charge transfer effects or phenomena such as the Stark effect \cite{XuePRB03}. This makes unnecessary the use of non-equilibrium Green's functions (NEGF) to calculate out-of-equilibrium charge densities and currents \cite{GarPRB12}, which further simplifies the problem.

\section{Results and discussion}

{\em BB nanogaps\label{sec:BB}}.$\--$ We study first the case of two clean surfaces or straight edges (BB). Ab-initio simulations of electronic transport in finite-width graphene layers terminated in straight zigzag graphene edges passivated with single hydrogen atoms show a featureless, almost flat and asymmetric transmission around the Fermi level ($E_\textrm{F}$). This is the type of transmission that is expected to appear in systems which have a constant number of bands around the Fermi level and which have states that couple very well to the bulk states (or are part of them) and which also couple across the gap. In case of graphene layers, both spin components of the transmission are slightly different and cross at a certain point near $E_\textrm{F}$. In order to reproduce these traits, we consider, as commented before, that the bulk state close to the border has a different on-site energy than that of the bulk states and some spin polarization due to the influence of magnetic impurities or magnetic edges. Notice that without such changes the transmission would be flat at $E_\textrm{F}$, much like the transmission of a perfect chain but smaller due to the weak coupling across the gap. Introducing a shift in the on-site energy creates a electron-hole asymmetry that makes the transmission have a finite slope at $E_\textrm{F}$. This happens because the transmission is higher at the energy of the on-site level and decreases away from it. The larger the shift towards negative (positive) energies, the bigger the negative (positive) slope. Since the on-site energy has also a exchange splitting, i.e. the energies for both spin components are different, the slopes for the spin-up and spin-down components turn out to be also different. This makes them cross at an energy point $\varepsilon_\uparrow+\varepsilon_\downarrow=2\varepsilon$.

\begin{figure}
\includegraphics[width=\columnwidth]{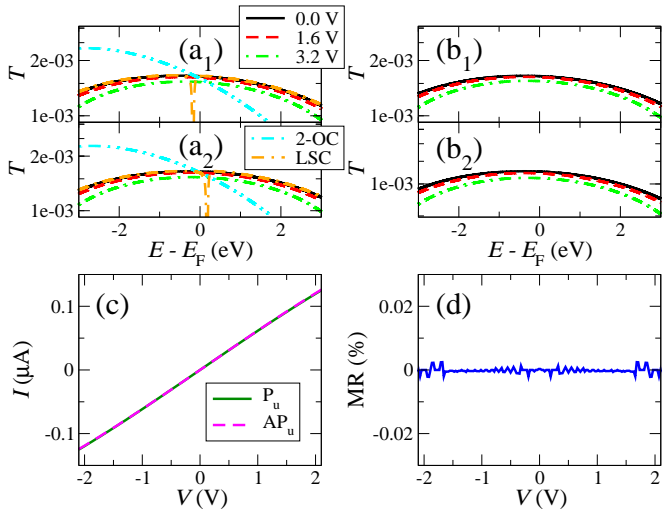}
\caption{\label{IV_bb}(Color online) Transmission at some positive voltages for the up (1) and down (2) spin channels of the parallel (a) and the antiparallel (b) configurations, current as a function of bias voltage for the spin-up channel of the parallel and the antiparallel configurations (c), and magnetoresistance ratio as a function of the bias voltage (d) for a system with uncoupled edge states (BB). Double-dotted dashed and double-dashed dotted lines in panels (a$_1$) and(a$_2$) represent spin up and down zero-bias transmissions for systems with second-order coupling across the gap (2-OC) or coupling to the edge states (LSC), respectively.}
\end{figure}

We compute the transmission function $T(E)$ for parallel (P) and antiparallel (AP) spin alignments of the magnetic impurities/edges. We use $t_{ij}=-3$ eV (the same values for both spins, unless otherwise stated), $\varepsilon=0$ eV, $t_{11}=\varepsilon_1=-0.1$ eV with an exchange splitting $\Delta=0.03$ eV ($\varepsilon_{1\uparrow(\downarrow)}=\varepsilon_1\varpm\Delta$) and $\gamma_{11}=-0.06$ eV. We obtain a featureless $T_\sigma(E)$ for both spin alignments in the energy window shown in Fig. (\ref{IV_bb}) due to the absence of coupling between the edge and bulk states. $T_\sigma(E)$ for the AP alignment is exactly the average of $T_\uparrow(E)$ and $T_\downarrow(E)$ for the P alignment, which produces a zero magnetoresistance ratio. This last quantity is defined as $\mathrm{MR}=(I^\mathrm{P}-I^\mathrm{AP})/(I^\mathrm{P}+I^\mathrm{AP})$ and is shown in panel (d). Fig. (\ref{IV_bb}) also shows the spin-resolved current-voltage characteristics in panel (c), which give a featureless differential conductance ($G$). This $IV$ curve, which reflects a typical ohmic behavior, could enable experimentalists to determine unambiguously the dependence of $G$ on the gap length $d$, measured as the distance between the atoms on both edges.

The bulk states could also couple across the gap to other bulk states located deeper in the electrodes. We call this coupling second-order coupling (2-OC). Relatively strong 2-OC, i.e. $\gamma_{22}=-0.03$ eV, can lead to more asymmetric transmission functions around $E_\textrm{F}$ due to  the emergence of antiresonances at higher energies (double-dotted dashed lines in panels (a$_1$) and (a$_2$) of Fig. (\ref{IV_bb})). On the other hand, the edge states can also weakly couple to the bulk states --without still coupling across the gap--, which we call localized-state coupling (LSC). This leads to antiresonances located at the positions of the on-site energies of the edge states (double-dashed dotted line in panels (a$_1$) and (a$_2$) of Fig. (\ref{IV_bb})). Both of these additional couplings do not change, however, the ohmic behavior.

\begin{figure}
\includegraphics[width=\columnwidth]{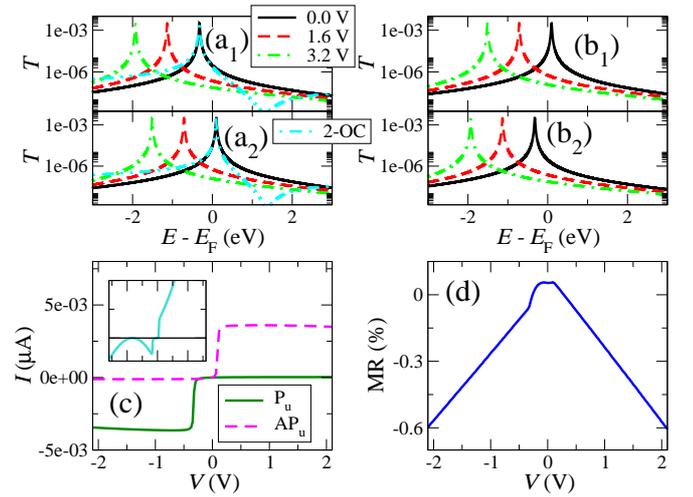}
\caption{\label{IV_bl}(Color online)
Transmission at some positive voltages for the up (1) and down (2) spin channels of the parallel (a) and the antiparallel (b) configurations, current as a function of bias voltage for the spin-up channel of the parallel and the antiparallel configurations (c), and magnetoresistance ratio as a function of the bias voltage (d) for a system with a clean surface/straight edge on one side and a impurity/wedge on the other (BL). Double-dotted dashed lines in panels (a$_1$) and(a$_2$) and the inset of (c) represent spin up and down zero-bias transmission and current for a system with second-order coupling across the gap (2-OC), respectively.}
\end{figure}

{\em BL nanogaps\label{sec:BL}}.$\--$ The results for BL nanogaps, where a clean surface/straight edge is in the left and a impurity/wedge is in the right, are shown in Fig. (\ref{IV_bl}). We use in this case $t_{\textrm{r},1\textrm{d}}=-0.17$ eV, $\epsilon=-0.1$ eV with an exchange splitting of 0.2 eV and $\gamma_{1\textrm{d}}=-4.8\times 10^{-3}$ eV. We keep the magnetic orientation of the left electrode in the up direction and switch upwards or downwards that of the right electrode. As a consequence, the transmission for both bias polarities turns out to be exactly the same, but with an opposite spin polarity, as can be seen in panels (a$_1$), (a$_2$), (b$_1$) and (b$_2$) of Fig. (\ref{IV_bl}). The transmission shows two sharp resonances located at energies $E=\epsilon_\sigma$ for both spin components. They correspond to localized states that couple asymmetrically to the bulk states on both sides of the gap.

The current-voltage characteristics, which agree with previous ab-initio simulations and experiments \cite{GarNs18}, can be understood by looking at the evolution of the transmission shown in panels (a) and (b) of Fig. (\ref{IV_bl}). When a positive bias is applied, both resonances shift down in energy according to $\epsilon_\sigma-eV/2$. In this case, only the spin-down resonance enters the bias window, suddenly increasing the spin-down current at a voltage $V=|\epsilon_{\downarrow}|/\textrm{e}$, while the spin-up current remains small. If the voltage increases further, the level stays inside the window, yielding then a constant current. When the bias reverses, the spin-up resonance enters the bias window and produces a sharp increase of the spin-up current at $V=-|\epsilon_{\uparrow}|/\textrm{e}$, while the spin-down component never enters the window and gives a low current signal. The current in both cases is fully spin-polarized, since one of the spin components is much bigger than the other for each bias polarity. This spin filtering effect would only be achievable, however, at low temperatures unless the magnetic anisotropy of these states is large enough to keep them aligned with a given magnetic configuration.

Changing some parameters can quantitatively alter the results, but the qualitative behavior remains the same. Increasing the coupling between the localized state and the bulk states in the same side, $t_{\textrm{r},1\textrm{d}}$, broadens the resonances but decreases their height because the the couplings of these states become more asymmetric. The broadening can induce a slight asymmetry in the $IV$ curve because the resonance that is closer to $E_\textrm{F}$ gives a larger contribution for the bias polarity for which it does not enter in the bias window and generates a larger current for that component. On a similar note, increasing the coupling across the gap, $\gamma_{1\textrm{d}}$, makes the couplings more symmetric and increases the height of the resonances. This enhances the total value of the current.

The asymmetry in the on-site energies of the levels produces also a clear rectification in the absolute bias range between the entrance of $\epsilon_{\downarrow}$ (for positive voltages) and the entrance of $\epsilon_{\uparrow}$ (for negative voltages) in the bias window, as can be seen in panel (c) of Fig. (\ref{IV_bl}). By changing some parameters it is possible to increase dramatically the rectification ratio (RR). For example, by using $\gamma_{1\mathrm{d}}=-7\times 10^{-3}$ eV, $t_{\mathrm{r},1}=-0.04$ eV, a splitting $\Delta=2.7$ eV and an on-site level centred at $\epsilon=-1.5$ eV, it is possible to achieve RR larger than $10^4$, which would give a rather good performance. There are basically three factors that affect the RR, in sequential order: (i) the position of the resonances relative to the Fermi level, i.e. the more asymmetric their positions (for instance one resonance close and the other away from the Fermi level) the higher the RR; (ii) the separation between the resonances, i.e. the more separated, the higher the RR; and (iii) the shape (width and height) of the resonances, i.e. the sharper the resonances, the higher the RR. Notice as well that, in order to maximize the RR, the nearer resonance to $E_\mathrm{F}$ should not be very close, so that the bias window starts covering a transmission as small as possible at the beginning.

Second order coupling of the edge states across the gap can also generate Fano antiresonances at high energies that alter the shape of the original resonances, as can be seen in panels (a$_1$) and (a$_2$) of Fig. (\ref{IV_bl}) (double-dotted dashed curve). If such coupling is relatively strong (e.g. $\gamma_{2\textrm{d}}=-0.01$ eV), which could happen if one of the layers is partly on top of the other, it can lead to non-trivial and highly asymmetric $IV$ curves (see the inset of panel (c)). This can also give rather high RR for a range of voltages due the suppression of transmission produced by the antiresonance for a certain bias polarity. In particular, for some combinations of parameters such as $\gamma_{1\mathrm{d}}=-4.8\times 10^{-3}$ eV, $\gamma_{2\mathrm{d}}=-0.01$ eV, $t_{\mathrm{r},1}=-0.03$ eV, a splitting $\Delta=0.8$ eV and an on-site level centred at $\epsilon=-0.5$ eV, it is again possible to achieve RR higher than $10^4$.

On the other hand, since both P and AP polarities give almost the same signal, the magnetoresistance turns out to be almost negligible, as shown in panel (d) of Fig. (\ref{IV_bl}). The finite MR values are due to the asymmetry introduced by the slightly polarized bulk states close to the surface.

\begin{figure}
\includegraphics[width=\columnwidth]{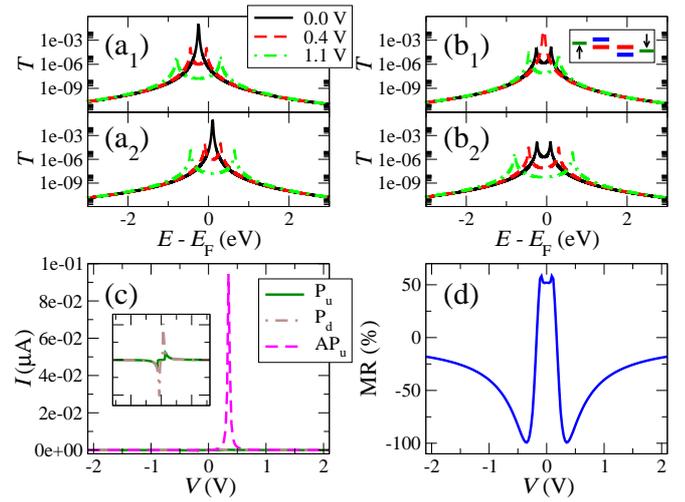}
\caption{\label{IV_ll}(Color online) Transmission at some positive voltages for the up (1) and down (2) spin channels of the parallel (a) and the antiparallel (b) configurations, current as a function of bias voltage for the spin-up channel of the parallel and the antiparallel configurations (c), and magnetoresistance ratio as a function of bias voltage (d) for a system with two impurities/wedges (LL). The inset of panel (c) shows a zoom-in of the current for the up and down spin channels of the parallel configuration.}
\end{figure}

{\em LL nanogaps\label{sec:LL}}.$\--$ We finally study the transport properties of LL nanogaps, where two impurities or wedges are facing each other. We use the same relevant parameters as in the previous cases, plus $\gamma_{\textrm{dd}}=-2\times 10^{-3}$ eV. We obtain similar zero-voltage transmission coefficients $T_\sigma(E)$ to those of a BL nanogap, at least in the P configuration, i.e. two spin-resolved resonances at energies $\epsilon_\sigma$ in the P configuration and two spin-unresolved and smaller resonances located at the same energies in the AP configuration. When a positive voltage is applied in the P configuration, the low-voltage current remains initially small because the levels dealign. The spin-down level on the left stays outside the integration window, while the spin-up level in the same electrode enters it at $V=|\epsilon_\uparrow|/\textrm{e}$, increasing the current. The rise is not very big, however, because both spin-up levels are dealigned. Increasing further the voltage decreases the current because both levels dealign even further, giving rise to a NDR feature. The $IV$ curve for the P alignment is shown in the inset of panel (c) in Fig. (\ref{IV_ll}).

For the AP alignment, the spin-up and -down levels switch at the right electrode. This leads to a rather different behavior as the voltage is raised from zero. Both spin-up levels move towards the integration window and towards each other. They eventually align, giving rise to a sharp increase of the transmission and the current. This is shown schematically in the inset of panel (b$_1$). As the voltage increases further, however, both levels dealign, decreasing the current. For the spin-down component, however, the situation reverses: the current stays small for positive biases because both spin-down levels at each side of the gap move away from each other and from the bias window. For negative voltages, the spin down levels enter the bias window, aligning with each other and later dealigning, which results in a large increase and decrease of the spin down current. The spin-resolved current-voltage for positive voltages is shown in the (c) panel of Fig. (\ref{IV_ll}). This produces then acute NDR features in the $IV$ curve for both bias polarities. The top of the peak at each bias polarity is reached at a voltage equal to $|V|=\left|\epsilon_{\uparrow}-\epsilon_{\downarrow}\right|/\textrm{e}$. The NDR signal is therefore determined by the location and coupling of the levels, which define the position and height of the peak, and its width.

The NDR peak is not affected by Fano resonances and is rather robust. Increasing the coupling with the bulk states in the same side, $t_{\textrm{l(r)},1\textrm{d}}$, broadens and reduces the NDR peak in the AP configuration, while enhances it in the P configuration. This is a consequence of the reduction of the height (specially in the P configuration) and the increase of the width of the resonances. On the other hand, increasing the coupling across the gap,  $\gamma_\textrm{dd}$, increases the NDR peak and maintains its aspect ratio. This is due to the increase of the height and width (specially in the P configuration) of the resonances. In general, the sharper the NDR peak, the better the operation of the device. This implies that $t_{\textrm{l(r)},1\textrm{d}}$ in particular should be as small as possible.

The large current peaks for the AP alignment, which are one order of magnitude larger than those resulting for the P alignment, give also rise to an almost perfect negative MR of $-100$ \% at the voltages of the peaks, as can be seen in panel (d) of Fig. (\ref{IV_ll}). Large positive MR values (more than 50 \%) can also be obtained at the voltages of the NDR peaks in the P alignment. Perfect spin rectification is also expected for this configuration. These spintronic features (MR and spin filtering/rectification) would only be possible, however, at low temperatures, unless the states have a relatively large magnetic anisotropy.

{\em Conclusions}.$\--$ We studied in this article novel electronic devices based on physical gaps, i.e. systems which only have two electrodes with certain atomic configurations or shapes. We discussed their transport properties and showed, using a tight-binding model, that, depending on the type of termination of both sides, it is possible to generate a plethora of electronic functionalities, which include ohmic behaviour, rectification, NDR and, in case of spin-polarized states, spin-filtering and magnetoresistance. In particular, we showed that clear electronic rectification, sharp NDR peaks, perfect spin rectification and very high magnetoresistance ratios, both positive and negative, can be achieved when one or both sides of the gap have (spin-polarized) localized states. This work defines a new paradigm of electronics that could lead to more functional and robust electronic systems.

\section*{Acknowledgements}
The research presented here was funded by the Spanish Ministerio de Econom\'{\i}a y Competitividad through the project FIS2015-63918-R and by the Spanish Ministerio de Ciencia, Innovaci\'on y Universidades through the project PGC2018-094783-B-I00. We acknowledge discussions with Amador Garc\'{\i}a-Fuente and Jaime Ferrer.

\section*{Author contributions}
VMGS did all the work.

\section*{Competing interests}
The authors declare no competing interests.

\section*{Data availability}
There are no restrictions on data availability.

\bibliographystyle{apsrev}
\bibliography{master}

\begin{thebibliography}{30}
\expandafter\ifx\csname natexlab\endcsname\relax\def\natexlab#1{#1}\fi
\expandafter\ifx\csname bibnamefont\endcsname\relax
  \def\bibnamefont#1{#1}\fi
\expandafter\ifx\csname bibfnamefont\endcsname\relax
  \def\bibfnamefont#1{#1}\fi
\expandafter\ifx\csname citenamefont\endcsname\relax
  \def\citenamefont#1{#1}\fi
\expandafter\ifx\csname url\endcsname\relax
  \def\url#1{\texttt{#1}}\fi
\expandafter\ifx\csname urlprefix\endcsname\relax\def\urlprefix{URL }\fi
\providecommand{\bibinfo}[2]{#2}
\providecommand{\eprint}[2][]{\url{#2}}

\bibitem[{\citenamefont{McCreery}(2004)}]{MccCM04}
\bibinfo{author}{\bibfnamefont{R.~L.} \bibnamefont{McCreery}},
  \bibinfo{journal}{Chem. Mater.} \textbf{\bibinfo{volume}{16}},
  \bibinfo{pages}{4477} (\bibinfo{year}{2004}).

\bibitem[{\citenamefont{Avouris et~al.}(2003)\citenamefont{Avouris,
  Appenzeller, Martel, and Wind}}]{AvoPIEEE03}
\bibinfo{author}{\bibfnamefont{P.}~\bibnamefont{Avouris}},
  \bibinfo{author}{\bibfnamefont{J.}~\bibnamefont{Appenzeller}},
  \bibinfo{author}{\bibfnamefont{R.}~\bibnamefont{Martel}}, \bibnamefont{and}
  \bibinfo{author}{\bibfnamefont{S.~J.} \bibnamefont{Wind}},
  \bibinfo{journal}{Proc. IEEE} \textbf{\bibinfo{volume}{91}},
  \bibinfo{pages}{1772} (\bibinfo{year}{2003}).

\bibitem[{\citenamefont{Palacios}(2012)}]{PalN12}
\bibinfo{author}{\bibfnamefont{T.}~\bibnamefont{Palacios}},
  \bibinfo{journal}{Nature} \textbf{\bibinfo{volume}{481}},
  \bibinfo{pages}{152} (\bibinfo{year}{2012}).

\bibitem[{\citenamefont{Osada and Sasaki}(2012)}]{OsaAM12}
\bibinfo{author}{\bibfnamefont{M.}~\bibnamefont{Osada}} \bibnamefont{and}
  \bibinfo{author}{\bibfnamefont{T.}~\bibnamefont{Sasaki}},
  \bibinfo{journal}{Adv. Mater.} \textbf{\bibinfo{volume}{24}},
  \bibinfo{pages}{210} (\bibinfo{year}{2012}).

\bibitem[{ITR(2015)}]{ITRS15}
\emph{\bibinfo{title}{International Technology Roadmap for Semiconductors
  (ITRS), 2015 Edition}} (\bibinfo{publisher}{Semiconductor Industry
  Association}, \bibinfo{year}{2015}).

\bibitem[{\citenamefont{Nguyen et~al.}(1989)\citenamefont{Nguyen, Cutler,
  Feuchtwang, Huang, Kuk, Silverman, Lucas, and Sullivan}}]{NguIEEETED89}
\bibinfo{author}{\bibfnamefont{H.~Q.} \bibnamefont{Nguyen}},
  \bibinfo{author}{\bibfnamefont{P.~H.} \bibnamefont{Cutler}},
  \bibinfo{author}{\bibfnamefont{T.~E.} \bibnamefont{Feuchtwang}},
  \bibinfo{author}{\bibfnamefont{Z.-H.} \bibnamefont{Huang}},
  \bibinfo{author}{\bibfnamefont{Y.}~\bibnamefont{Kuk}},
  \bibinfo{author}{\bibfnamefont{P.~J.} \bibnamefont{Silverman}},
  \bibinfo{author}{\bibfnamefont{A.~A.} \bibnamefont{Lucas}}, \bibnamefont{and}
  \bibinfo{author}{\bibfnamefont{T.~E.} \bibnamefont{Sullivan}},
  \bibinfo{journal}{IEEE Transac. Elec. Devices} \textbf{\bibinfo{volume}{36}},
  \bibinfo{pages}{2671} (\bibinfo{year}{1989}).

\bibitem[{\citenamefont{Joachim and Gimzewski}(1997)}]{JoaCPL97}
\bibinfo{author}{\bibfnamefont{C.}~\bibnamefont{Joachim}} \bibnamefont{and}
  \bibinfo{author}{\bibfnamefont{J.~K.} \bibnamefont{Gimzewski}},
  \bibinfo{journal}{Chem. Phys. Lett.} \textbf{\bibinfo{volume}{265}},
  \bibinfo{pages}{353} (\bibinfo{year}{1997}).

\bibitem[{\citenamefont{Lyo and Avouris}(1989)}]{LyoS89}
\bibinfo{author}{\bibfnamefont{I.-W.} \bibnamefont{Lyo}} \bibnamefont{and}
  \bibinfo{author}{\bibfnamefont{P.}~\bibnamefont{Avouris}},
  \bibinfo{journal}{Science} \textbf{\bibinfo{volume}{245}},
  \bibinfo{pages}{1369} (\bibinfo{year}{1989}).

\bibitem[{\citenamefont{Chen et~al.}(2000)\citenamefont{Chen, Wang, Reed,
  Rawlett, Price, and Tour}}]{CheAPL00}
\bibinfo{author}{\bibfnamefont{J.}~\bibnamefont{Chen}},
  \bibinfo{author}{\bibfnamefont{W.}~\bibnamefont{Wang}},
  \bibinfo{author}{\bibfnamefont{M.~A.} \bibnamefont{Reed}},
  \bibinfo{author}{\bibfnamefont{A.~M.} \bibnamefont{Rawlett}},
  \bibinfo{author}{\bibfnamefont{D.~W.} \bibnamefont{Price}}, \bibnamefont{and}
  \bibinfo{author}{\bibfnamefont{J.~M.} \bibnamefont{Tour}},
  \bibinfo{journal}{Applied Physics Letters} \textbf{\bibinfo{volume}{77}},
  \bibinfo{pages}{1224} (\bibinfo{year}{2000}).

\bibitem[{\citenamefont{G\"ohler et~al.}(2011)\citenamefont{G\"ohler,
  Hamelbeck, Markus, Kettner, Hanne, Vager, Naaman, and Zacharias}}]{GohS211}
\bibinfo{author}{\bibfnamefont{B.}~\bibnamefont{G\"ohler}},
  \bibinfo{author}{\bibfnamefont{V.}~\bibnamefont{Hamelbeck}},
  \bibinfo{author}{\bibfnamefont{T.~Z.} \bibnamefont{Markus}},
  \bibinfo{author}{\bibfnamefont{M.}~\bibnamefont{Kettner}},
  \bibinfo{author}{\bibfnamefont{G.~F.} \bibnamefont{Hanne}},
  \bibinfo{author}{\bibfnamefont{Z.}~\bibnamefont{Vager}},
  \bibinfo{author}{\bibfnamefont{R.}~\bibnamefont{Naaman}}, \bibnamefont{and}
  \bibinfo{author}{\bibfnamefont{H.}~\bibnamefont{Zacharias}},
  \bibinfo{journal}{Science} \textbf{\bibinfo{volume}{331}},
  \bibinfo{pages}{894} (\bibinfo{year}{2011}).

\bibitem[{\citenamefont{Tsukagoshi et~al.}(1999)\citenamefont{Tsukagoshi,
  Alphenaar, and Ago}}]{TsuN99}
\bibinfo{author}{\bibfnamefont{K.}~\bibnamefont{Tsukagoshi}},
  \bibinfo{author}{\bibfnamefont{B.~W.} \bibnamefont{Alphenaar}},
  \bibnamefont{and} \bibinfo{author}{\bibfnamefont{H.}~\bibnamefont{Ago}},
  \bibinfo{journal}{Nature} \textbf{\bibinfo{volume}{401}},
  \bibinfo{pages}{572} (\bibinfo{year}{1999}).

\bibitem[{\citenamefont{Garc\'{\i}a et~al.}(1999)\citenamefont{Garc\'{\i}a,
  Munoz, and Zhao}}]{GarPRL99}
\bibinfo{author}{\bibfnamefont{N.}~\bibnamefont{Garc\'{\i}a}},
  \bibinfo{author}{\bibfnamefont{M.}~\bibnamefont{Munoz}}, \bibnamefont{and}
  \bibinfo{author}{\bibfnamefont{Y.-W.} \bibnamefont{Zhao}},
  \bibinfo{journal}{Phys. Rev. Lett.} \textbf{\bibinfo{volume}{82}},
  \bibinfo{pages}{2923} (\bibinfo{year}{1999}).

\bibitem[{\citenamefont{L\"ortscher}(2013)}]{LorNN13}
\bibinfo{author}{\bibfnamefont{E.}~\bibnamefont{L\"ortscher}},
  \bibinfo{journal}{Nat. Nanotech.} \textbf{\bibinfo{volume}{8}},
  \bibinfo{pages}{381} (\bibinfo{year}{2013}).

\bibitem[{\citenamefont{Moore}(1965)}]{MorE65}
\bibinfo{author}{\bibfnamefont{G.~E.} \bibnamefont{Moore}},
  \bibinfo{journal}{Electronics} \textbf{\bibinfo{volume}{38}},
  \bibinfo{pages}{114} (\bibinfo{year}{1965}).

\bibitem[{\citenamefont{Topaloglu}(2015)}]{Top15Mo}
\bibinfo{author}{\bibfnamefont{R.~O.} \bibnamefont{Topaloglu}},
  \emph{\bibinfo{title}{More than Moore Technologies for Next Generation
  Computer Design}} (\bibinfo{publisher}{Springer}, \bibinfo{year}{2015}).

\bibitem[{\citenamefont{Koleini et~al.}(2007)\citenamefont{Koleini, Paulsson,
  and Brandbyge}}]{KolPRL07}
\bibinfo{author}{\bibfnamefont{M.}~\bibnamefont{Koleini}},
  \bibinfo{author}{\bibfnamefont{M.}~\bibnamefont{Paulsson}}, \bibnamefont{and}
  \bibinfo{author}{\bibfnamefont{M.}~\bibnamefont{Brandbyge}},
  \bibinfo{journal}{Phys. Rev. Lett.} \textbf{\bibinfo{volume}{98}},
  \bibinfo{pages}{197202} (\bibinfo{year}{2007}).

\bibitem[{\citenamefont{Prins et~al.}(2011)\citenamefont{Prins, Barreiro,
  Ruitenberg, Seldenthius, Aliaga-Alcalde, Vandersypen, and van~der
  Zant}}]{PriNL11}
\bibinfo{author}{\bibfnamefont{F.}~\bibnamefont{Prins}},
  \bibinfo{author}{\bibfnamefont{A.}~\bibnamefont{Barreiro}},
  \bibinfo{author}{\bibfnamefont{J.~M.} \bibnamefont{Ruitenberg}},
  \bibinfo{author}{\bibfnamefont{J.~S.} \bibnamefont{Seldenthius}},
  \bibinfo{author}{\bibfnamefont{N.}~\bibnamefont{Aliaga-Alcalde}},
  \bibinfo{author}{\bibfnamefont{L.~M.~K.} \bibnamefont{Vandersypen}},
  \bibnamefont{and} \bibinfo{author}{\bibfnamefont{H.~S.~J.}
  \bibnamefont{van~der Zant}}, \bibinfo{journal}{Nano Lett.}
  \textbf{\bibinfo{volume}{11}}, \bibinfo{pages}{4607} (\bibinfo{year}{2011}).

\bibitem[{\citenamefont{Martin et~al.}(2008)\citenamefont{Martin, Ding,
  S{\o}rensen, Bj{\o}rnholm, van Ruitenbeek, and van~der Zant}}]{MarJACS08}
\bibinfo{author}{\bibfnamefont{C.~A.} \bibnamefont{Martin}},
  \bibinfo{author}{\bibfnamefont{D.}~\bibnamefont{Ding}},
  \bibinfo{author}{\bibfnamefont{J.~K.} \bibnamefont{S{\o}rensen}},
  \bibinfo{author}{\bibfnamefont{T.}~\bibnamefont{Bj{\o}rnholm}},
  \bibinfo{author}{\bibfnamefont{J.~M.} \bibnamefont{van Ruitenbeek}},
  \bibnamefont{and} \bibinfo{author}{\bibfnamefont{H.~S.~J.}
  \bibnamefont{van~der Zant}}, \bibinfo{journal}{J. Am. Chem. Soc.}
  \textbf{\bibinfo{volume}{130}}, \bibinfo{pages}{13198}
  (\bibinfo{year}{2008}).

\bibitem[{\citenamefont{Garc\'{\i}a-Su\'arez
  et~al.}(2013)\citenamefont{Garc\'{\i}a-Su\'arez, Ferrad\'as, Carrascal, and
  Ferrer}}]{GarPRB13}
\bibinfo{author}{\bibfnamefont{V.~M.} \bibnamefont{Garc\'{\i}a-Su\'arez}},
  \bibinfo{author}{\bibfnamefont{R.}~\bibnamefont{Ferrad\'as}},
  \bibinfo{author}{\bibfnamefont{D.}~\bibnamefont{Carrascal}},
  \bibnamefont{and} \bibinfo{author}{\bibfnamefont{J.}~\bibnamefont{Ferrer}},
  \bibinfo{journal}{Phys. Rev. B} \textbf{\bibinfo{volume}{87}},
  \bibinfo{pages}{235425} (\bibinfo{year}{2013}).

\bibitem[{\citenamefont{de~Abajo et~al.}(2010)\citenamefont{de~Abajo, Cord\'on,
  Corso, Schiller, and Ortega}}]{GarN10}
\bibinfo{author}{\bibfnamefont{F.~J.~G.} \bibnamefont{de~Abajo}},
  \bibinfo{author}{\bibfnamefont{J.}~\bibnamefont{Cord\'on}},
  \bibinfo{author}{\bibfnamefont{M.}~\bibnamefont{Corso}},
  \bibinfo{author}{\bibfnamefont{F.}~\bibnamefont{Schiller}}, \bibnamefont{and}
  \bibinfo{author}{\bibfnamefont{J.~E.} \bibnamefont{Ortega}},
  \bibinfo{journal}{Nanoscale} \textbf{\bibinfo{volume}{2}},
  \bibinfo{pages}{717} (\bibinfo{year}{2010}).

\bibitem[{\citenamefont{Xiu et~al.}(2011)\citenamefont{Xiu, He, Wang, Cheng,
  Chang, Lang, Huang, Kou, Zhou, Jiang et~al.}}]{XiuNN11}
\bibinfo{author}{\bibfnamefont{F.}~\bibnamefont{Xiu}},
  \bibinfo{author}{\bibfnamefont{L.}~\bibnamefont{He}},
  \bibinfo{author}{\bibfnamefont{Y.}~\bibnamefont{Wang}},
  \bibinfo{author}{\bibfnamefont{L.}~\bibnamefont{Cheng}},
  \bibinfo{author}{\bibfnamefont{L.-T.} \bibnamefont{Chang}},
  \bibinfo{author}{\bibfnamefont{M.}~\bibnamefont{Lang}},
  \bibinfo{author}{\bibfnamefont{G.}~\bibnamefont{Huang}},
  \bibinfo{author}{\bibfnamefont{X.}~\bibnamefont{Kou}},
  \bibinfo{author}{\bibfnamefont{Y.}~\bibnamefont{Zhou}},
  \bibinfo{author}{\bibfnamefont{X.}~\bibnamefont{Jiang}},
  \bibnamefont{et~al.}, \bibinfo{journal}{Nat. Nanotech.}
  \textbf{\bibinfo{volume}{6}}, \bibinfo{pages}{216} (\bibinfo{year}{2011}).

\bibitem[{\citenamefont{Nirantar et~al.}(2018)\citenamefont{Nirantar, Ahmed,
  Ren, Gutruf, Xu, Bhaskaran, Walia, and Sriram}}]{NirNL18}
\bibinfo{author}{\bibfnamefont{S.}~\bibnamefont{Nirantar}},
  \bibinfo{author}{\bibfnamefont{T.}~\bibnamefont{Ahmed}},
  \bibinfo{author}{\bibfnamefont{G.}~\bibnamefont{Ren}},
  \bibinfo{author}{\bibfnamefont{P.}~\bibnamefont{Gutruf}},
  \bibinfo{author}{\bibfnamefont{C.}~\bibnamefont{Xu}},
  \bibinfo{author}{\bibfnamefont{M.}~\bibnamefont{Bhaskaran}},
  \bibinfo{author}{\bibfnamefont{S.}~\bibnamefont{Walia}}, \bibnamefont{and}
  \bibinfo{author}{\bibfnamefont{S.}~\bibnamefont{Sriram}},
  \bibinfo{journal}{Nano Lett.} \textbf{\bibinfo{volume}{18}},
  \bibinfo{pages}{7478} (\bibinfo{year}{2018}).

\bibitem[{\citenamefont{Katkov and Osipov}(2017)}]{KatJVSTB17}
\bibinfo{author}{\bibfnamefont{V.~L.} \bibnamefont{Katkov}} \bibnamefont{and}
  \bibinfo{author}{\bibfnamefont{V.~A.} \bibnamefont{Osipov}},
  \bibinfo{journal}{J. Vac. Sci. Technol. B} \textbf{\bibinfo{volume}{35}},
  \bibinfo{pages}{050801} (\bibinfo{year}{2017}).

\bibitem[{\citenamefont{Garc\'{\i}a-Su\'arez
  et~al.}(2018)\citenamefont{Garc\'{\i}a-Su\'arez, Garc\'{\i}a-Fuente,
  Carrascal, Burzur\'{\i}, Koole, van~der Zant, El~Abbassi, Calame, and
  Ferrer}}]{GarNs18}
\bibinfo{author}{\bibfnamefont{V.~M.} \bibnamefont{Garc\'{\i}a-Su\'arez}},
  \bibinfo{author}{\bibfnamefont{A.}~\bibnamefont{Garc\'{\i}a-Fuente}},
  \bibinfo{author}{\bibfnamefont{D.~J.} \bibnamefont{Carrascal}},
  \bibinfo{author}{\bibfnamefont{E.}~\bibnamefont{Burzur\'{\i}}},
  \bibinfo{author}{\bibfnamefont{M.}~\bibnamefont{Koole}},
  \bibinfo{author}{\bibfnamefont{H.~S.~J.} \bibnamefont{van~der Zant}},
  \bibinfo{author}{\bibfnamefont{M.}~\bibnamefont{El~Abbassi}},
  \bibinfo{author}{\bibfnamefont{M.}~\bibnamefont{Calame}}, \bibnamefont{and}
  \bibinfo{author}{\bibfnamefont{J.}~\bibnamefont{Ferrer}},
  \bibinfo{journal}{Nanoscale} \textbf{\bibinfo{volume}{10}},
  \bibinfo{pages}{18169} (\bibinfo{year}{2018}).

\bibitem[{\citenamefont{Caneva et~al.}(2018)\citenamefont{Caneva, Gehring,
  Garc\'{\i}a-Su\'arez, Garc\'{\i}a-Fuente, Stefani, Olavarria-Contreras,
  Ferrer, Dekker, and van~der Zant}}]{CanNN18}
\bibinfo{author}{\bibfnamefont{S.}~\bibnamefont{Caneva}},
  \bibinfo{author}{\bibfnamefont{P.}~\bibnamefont{Gehring}},
  \bibinfo{author}{\bibfnamefont{V.~M.} \bibnamefont{Garc\'{\i}a-Su\'arez}},
  \bibinfo{author}{\bibfnamefont{A.}~\bibnamefont{Garc\'{\i}a-Fuente}},
  \bibinfo{author}{\bibfnamefont{D.}~\bibnamefont{Stefani}},
  \bibinfo{author}{\bibfnamefont{I.~J.} \bibnamefont{Olavarria-Contreras}},
  \bibinfo{author}{\bibfnamefont{J.}~\bibnamefont{Ferrer}},
  \bibinfo{author}{\bibfnamefont{C.}~\bibnamefont{Dekker}}, \bibnamefont{and}
  \bibinfo{author}{\bibfnamefont{H.~S.~J.} \bibnamefont{van~der Zant}},
  \bibinfo{journal}{Nat. Nanotech.} \textbf{\bibinfo{volume}{13}},
  \bibinfo{pages}{1126} (\bibinfo{year}{2018}).

\bibitem[{\citenamefont{Mart\'{\i}n et~al.}(2009)\citenamefont{Mart\'{\i}n,
  Manrique, Garc\'{\i}a-Su\'arez, Haiss, Higgins, Lambert, and
  Nichols}}]{MarNt09}
\bibinfo{author}{\bibfnamefont{S.}~\bibnamefont{Mart\'{\i}n}},
  \bibinfo{author}{\bibfnamefont{D.~Z.} \bibnamefont{Manrique}},
  \bibinfo{author}{\bibfnamefont{V.~M.} \bibnamefont{Garc\'{\i}a-Su\'arez}},
  \bibinfo{author}{\bibfnamefont{W.}~\bibnamefont{Haiss}},
  \bibinfo{author}{\bibfnamefont{S.~J.} \bibnamefont{Higgins}},
  \bibinfo{author}{\bibfnamefont{C.~J.} \bibnamefont{Lambert}},
  \bibnamefont{and} \bibinfo{author}{\bibfnamefont{R.~J.}
  \bibnamefont{Nichols}}, \bibinfo{journal}{Nanotechnology}
  \textbf{\bibinfo{volume}{20}}, \bibinfo{pages}{125203}
  (\bibinfo{year}{2009}).

\bibitem[{\citenamefont{Sparks et~al.}(2011)\citenamefont{Sparks,
  Garc\'{\i}a-Su\'arez, Manrique, and Lambert}}]{SpaPRB11}
\bibinfo{author}{\bibfnamefont{R.~E.} \bibnamefont{Sparks}},
  \bibinfo{author}{\bibfnamefont{V.~M.} \bibnamefont{Garc\'{\i}a-Su\'arez}},
  \bibinfo{author}{\bibfnamefont{D.~Z.} \bibnamefont{Manrique}},
  \bibnamefont{and} \bibinfo{author}{\bibfnamefont{C.~J.}
  \bibnamefont{Lambert}}, \bibinfo{journal}{Phys. Rev. B}
  \textbf{\bibinfo{volume}{83}}, \bibinfo{pages}{075437}
  (\bibinfo{year}{2011}).

\bibitem[{\citenamefont{Carrascal et~al.}(2012)\citenamefont{Carrascal,
  Garc\'{\i}a-Su\'arez, and Ferrer}}]{CarPRB12}
\bibinfo{author}{\bibfnamefont{D.}~\bibnamefont{Carrascal}},
  \bibinfo{author}{\bibfnamefont{V.~M.} \bibnamefont{Garc\'{\i}a-Su\'arez}},
  \bibnamefont{and} \bibinfo{author}{\bibfnamefont{J.}~\bibnamefont{Ferrer}},
  \bibinfo{journal}{Phys. Rev. B} \textbf{\bibinfo{volume}{85}},
  \bibinfo{pages}{195434} (\bibinfo{year}{2012}).

\bibitem[{\citenamefont{Xue and Ratner}(2003)}]{XuePRB03}
\bibinfo{author}{\bibfnamefont{Y.}~\bibnamefont{Xue}} \bibnamefont{and}
  \bibinfo{author}{\bibfnamefont{M.~A.} \bibnamefont{Ratner}},
  \bibinfo{journal}{Phys. Rev. B} \textbf{\bibinfo{volume}{68}},
  \bibinfo{pages}{115406} (\bibinfo{year}{2003}).

\bibitem[{\citenamefont{Garc\'{\i}a-Su\'arez and Ferrer}(2012)}]{GarPRB12}
\bibinfo{author}{\bibfnamefont{V.~M.} \bibnamefont{Garc\'{\i}a-Su\'arez}}
  \bibnamefont{and} \bibinfo{author}{\bibfnamefont{J.}~\bibnamefont{Ferrer}},
  \bibinfo{journal}{Phys. Rev. B} \textbf{\bibinfo{volume}{86}},
  \bibinfo{pages}{125446} (\bibinfo{year}{2012}).

\end{thebibliography}

\end{document}